\documentclass{ws-ijmpa}

\begin{document}

\markboth{Steven R. Elliott}
{Experiments for Neutrinoless Double-Beta Decay}

%
\catchline{}{}{}{}{}
%


\def\nuc#1#2{${}^{#1}$#2}
\def\mee{$\langle m_{\beta\beta} \rangle$}
\def\mnu{$\langle m_{\nu} \rangle$}
\def\mmod{$\| \langle m_{ee} \rangle \|$}
\def\mb{$\langle m_{\beta} \rangle$}
\def\BBz{$\beta\beta(0\nu)$}
\def\BBt{$\beta\beta(2\nu)$}
\def\BB{$\beta\beta$}
\def\Mz{$|M_{0\nu}|$}
\def\Mt{$|M_{2\nu}|$}
\def\Tz{$T^{0\nu}_{1/2}$}
\def\Tt{$T^{2\nu}_{1/2}$}
\def\Tc{$T^{0\nu\,\chi}_{1/2}$}
\def\today{\space\number\day\space\ifcase\month\or January\or February\or
  March\or April\or May\or June\or July\or August\or September\or October\or
  November\or December\fi\space\number\year}

\input psfig.sty

\title{Experiments for Neutrinoless Double-Beta Decay}

\author{\footnotesize Steven R. Elliott}

\address{Physics Division, Los Alamos National Laboratory\\
MS H803, P-23, Los Alamos, NM, 87545, USA}

\maketitle

\pub{Received (Day Month Year)}{Revised (Day Month Year)}

\begin{abstract}
The recent neutrino oscillation experimental results indicate that at
least one neutrino has a mass greater than 50 meV. The next generation 
of double-beta decay experiments will very likely have a sensitivity
to an effective Majorana neutrino mass below this target. Therefore this is a very 
exciting time for this field of research as even null results from
these experiments have the potential to elucidate the nature of the 
neutrino. 

\end{abstract}

\section{Introduction}	
Double-beta decay (\BB) is a nuclear process where the nucleus increases in charge by 2 while 
emitting 2 electrons. Two-neutrino double-beta decay (\BBt) is an allowed second order
weak process that also emits two neutrinos. Zero-neutrino double-beta decay (\BBz) can only occur
if neutrinos are massive particles that are self conjugate (i.e. massive Majorana particles).

The recent experimental results from  atmospheric\cite{SKatm01,Kam94,IMB92,SOU99}, 
solar\cite{chlorine,Sage,Gallex,SNO,SKsol01},
and reactor\cite{Chooz99,Palo01,KamLAND} neutrino sources indicate neutrinos have mass and that
they mix. In particular, at least one neutrino must have a mass greater than about 50 meV. 
This very exciting result has greatly renewed the interest in \BBz\
 because the next generation experiments aim to reach this target.
 In fact of all the experimental techniques to search for neutrino mass, only \BBz\ experiments can reach
 an absolute mass value near the 50-meV mass scale indicated by the oscillation 
experiments. Furthermore, only \BBz\ has
the potential to determine whether the neutrino is its own anti-particle.  This exciting situation
and the present experimental and theoretical status has been described in a recent review\cite{ELL02} and the many references therein. 

Although any
observation of \BBz\ has always been an exciting possibility, in the not-to-recent past, limits 
on the Majorana mass from the non-observation of \BBz\ could be interpreted in two distinct 
ways. Perhaps the neutrino mass was just smaller than the limit. Or, since \BBz\ is only sensitive
to Majorana neutrinos, it could be that the neutrino
was a Dirac particle that has a mass greater than the limit. Now, however, there is a 
known minimum mass scale for the neutrino.  A null result below this mass scale 
 could eliminate presently-possible mass spectra from having a Majorana component. 
Therefore such a result would be a non-trivial constraint on neutrino properties.  

The oscillation results convincingly indicate two mass-difference scales:
One at $\delta m_{atm}^2 \sim (50$ meV$)^2$ at approximately maximal mixing and one at
$\delta m_{sol}^2 \sim (7$ meV$)^2$ at a mixing angle near 30 degrees. These results 
indicate that neutrinos do have mass but do not give an indication of the absolute neutrino
mass scale or its CP nature. Furthermore, we have little data on the hierarchy of the
neutrino masses. That is we do not know if $\nu_e$ is predominately composed of the lightest
mass eigenstate or not. \BBz\ can provide data to address these neutrino properties, but
because of  possible non-trivial Majorana phases in the mixing matrix, we will also
require input from kinematic neutrino mass experiments such as Katrin\cite{Katrin}. Therefore, all three 
types of experiments, oscillation, \BBz, and kinematic, are required to completely understand the 
neutrino. Note that indications of a
third mass-difference scale near 1 eV$^2$ from the LSND experiment\cite{LSND} are as yet unconfirmed.
Although I have chosen to ignore that result in this simple discussion, the large mass
scale would only add to the interest in \BB\ experiments.

The decay rate ($\Gamma$) of double beta decay is proportional to the square of the effective
Majorana neutrino mass (\mee), an easily calculable phase space factor ($G$), and a 
difficult-to-calculate nuclear matrix element ($M_{0\nu}$); 

\begin{equation}
\Gamma = G M_{0\nu} \langle m_{\beta\beta} \rangle^2\mbox{.}
\end{equation}

The value for \mee\ in turn, depends on the values of the individual neutrino mass eigenstates ($m_i$), the
mixing matrix elements ($U_{ei}$) and the Majorana phases ($\alpha_i$);

\begin{equation}
\langle m_{\beta\beta} \rangle^2 = \left| \sum_i^N  |U_{e i}|^2 e^{\alpha_i} m_i \right|^2\mbox{.}
\end{equation}

In the case of CP conservation $e^{\alpha_i} = \pm 1$. If we take $U_{e3} = 0$, the atmospheric mixing to
be maximal, and the solar mixing angle to be about 30 degrees ($U_{e1}^2 = 0.25$, $U_{e2}^2 = 0.75$), 
one can simplify this equation for \mee\ and get a
feeling as to the sensitivity that \BBz\ has to the absolute mass scale.

\begin{equation}
\langle m_{\beta\beta} \rangle \sim  |U_{e 1}|^2 m_1 \pm |U_{e 2}|^2 \sqrt{m_1^2 + \delta m_{12}^2} \mbox{.}
\end{equation}

For Majorana neutrinos in the normal hierarchy, one finds \mee\ values greater than several 
meV and in the inverse hierarchy one 
finds values greater than several tens of meV even when the smallest mass eigenstate is near zero. Simply stated,
$\sqrt{\delta m_{atm}^2}$ is a very enticing target for \BBz\ experiments and the recent
excitement in \BBz\ is that the next generation of experiments will be sensitive to \mee\ near this target.

\section{Backgrounds to Double-Beta Decay}

To observe \BBz, one must reduce backgrounds so that an ultra-low-rate peak can emerge from a continuum of other processes.
To reach the desired 50-meV goal, significant background improvements 
 are needed. The backgrounds that currently afflict \BBz\ experiments can be classified into 3 groups: \BBt, natural or 
man-made radioactivities, and cosmogenicly produced radioactivities.

\subsection{Two-Neutrino Double-Beta Decay as a Background}
Unlike \BBz, the two electrons share the available energy with two neutrinos in the \BBt\ process. Thus their
sum energy spectrum is a distribution up to the endpoint. (See Fig. 1.) This spectrum is very
steeply falling and, in principle, the region of interest for \BBz\ should be free of such events. However,
the finite resolution of any detector can result in \BBt\ events polluting the \BBz\ region. An approximate 
expression\cite{ELL02} for the \mee\ sensitivity limit due to this background can be written

\begin{equation}
\langle m_{\beta\beta} \rangle^2  \sim  
\frac{7Q \delta^6}{m_e} \frac{G_{2\nu}}{G_{0\nu}} \frac{|M_{2\nu}|^2}{|M_{0\nu}|^2}
\end{equation}

\noindent where $Q$ is the Q-value, $\delta$ is the width of the region of interest divided by $Q$,
$m_e$ is the electron mass, $G_{2\nu}$ is the \BBt\ phase space factor, 
$G_{0\nu}$ is the \BBz\ phase space factor, \Mt\ is the \BBt\ matrix element and
\Mz\ is the \BBz\ matrix element.

\begin{figure}[htb]
\begin{center}
\epsfysize=3.5in \epsfbox{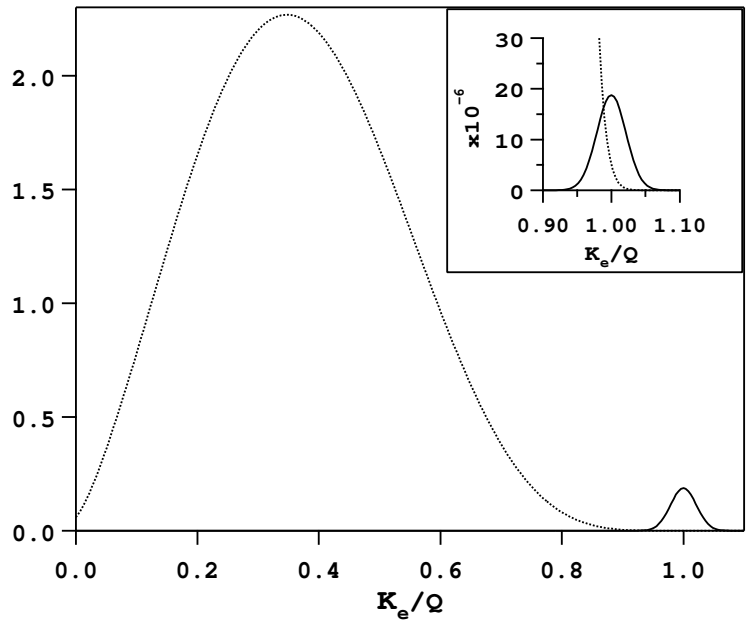} 
\vspace*{8pt}
\caption{Illustration of the spectra of the sum of the electron kinetic energies 
K$_e$ (Q is the endpoint) for the 
\BBt\, normalized to 1 (dotted curve)
and \BBz\ decays (solid curve). The \BBz\ spectrum is normalized to $10^{-2}$ 
($10^{-6}$ in the figure inset).
All spectra are convolved with
an energy resolution of 5\%, representative of several
experiments.  However, some experiments, notably Ge diodes and Te bolometers, have  
a much better energy resolution ($\approx0.2$\%). }
\end{center}
\label{twonuspect}
\end{figure}

Although this equation does not include the improvement one might get from exploiting other kinematic
measurements such as the opening angle or the individual electron energies, it does give an estimate of
the resolution requirements. This equation is plotted for a few isotopes in Fig. 2 and one sees that a resolution better a few percent will
suffice for the 50-meV goal. However, if other backgrounds are present and create a continuum through the 
region of interest, better resolution improves the experimental sensitivity.

\begin{figure}[htb]
\begin{center}
\epsfysize=4.0in \epsfbox{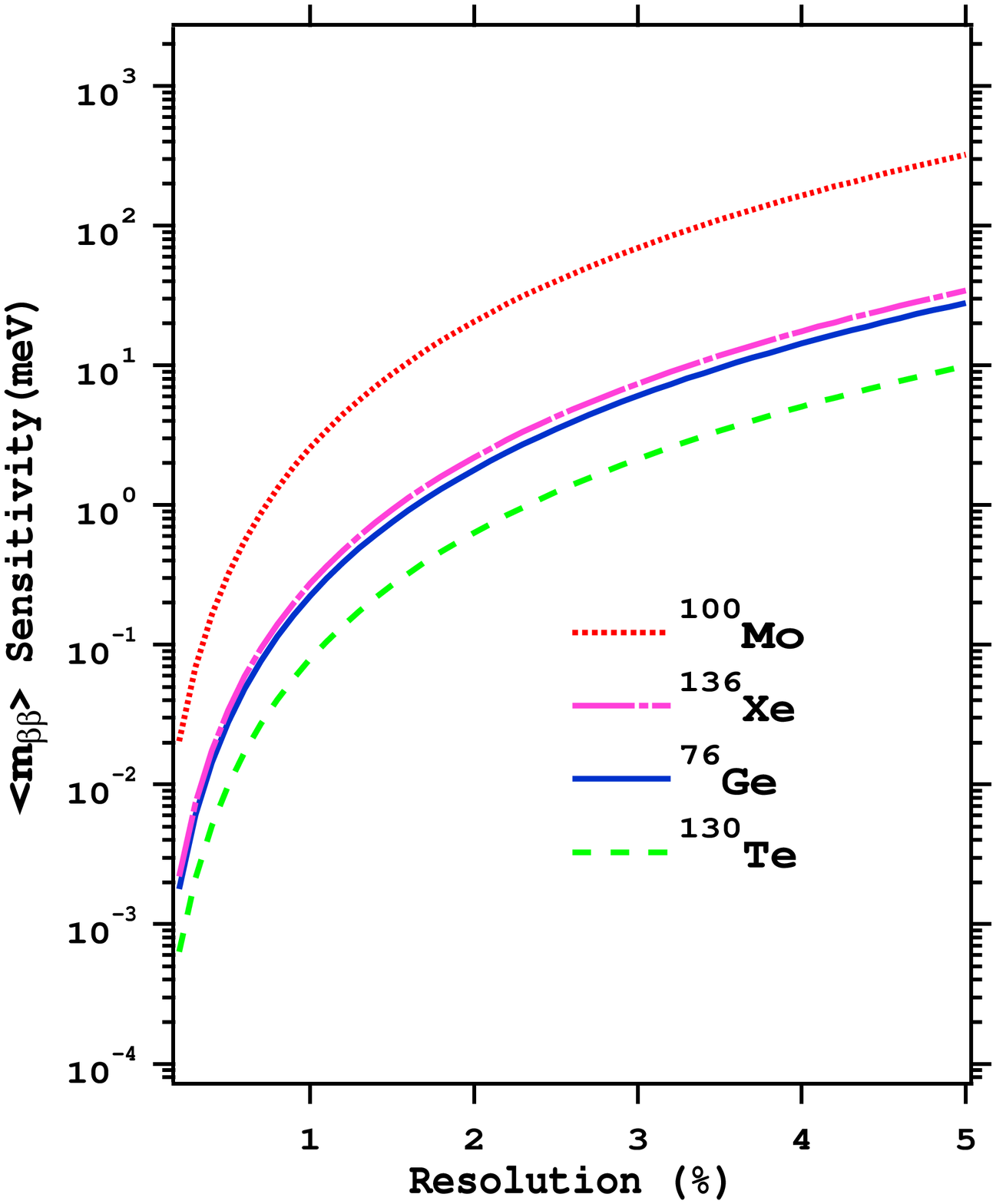} 
\vspace*{8pt}
\caption{The limit on the mass sensitivity as a function of sum energy resolution for a few isotopes.}
\end{center}
\label{twonuback}
\end{figure}

\subsection{Natural and Man-made Radioactivity}
Many \BB\ experiments also serve as dark matter searches. Those searches look for the low-energy recoils resulting
from elastic scattering of Weakly Interacting Massive Particles (WIMPS). The potential background for those
searches is more varied than for \BBz. Because the \BBz\ endpoint is typically a few MeV,
many natural radioactivities simply contribute too little energy to pollute that region of interest. 

The most important naturally occurring isotopes that are potential backgrounds for \BBz\ are
 $^{208}$Tl and $^{214}$Bi. These have large Q-values and can pollute the region of interest of almost all \BB\ isotopes.
They are members of the natural Th and U decay chains and thus common in the environment. Furthermore,
they are daughters of the gaseous Rn isotopes, which are very mobile. The Th and U half-lives,
 ($\approx 10^{10}$ y), are much shorter than the required $10^{26}$ to $10^{27}$-y sensitivity for the 50-meV goal. Therefore
even a tiny amount of these activities are a significant problem. Over the past 60 years, experimentalists have
made great progress in identifying materials that are very low in Th and U. By building their experiments from 
this limited palate of materials, these activities have been greatly reduced. Improved purification 
techniques have also helped
eliminate these backgrounds. 

Radon is a special problem because its a gas that emanates from U and Th containing compounds and diffuses through many 
materials also.  Experimenters must ensure that the detector volume is kept free of Rn. In many cases a careful
flushing of the atmosphere near the inner volume with boil-off gas 
from liquid nitrogen sufficiently reduces the Rn. At LN temperatures, Rn
is frozen out and therefore the boil-off gas is mostly free of Rn.

There are techniques to tag Tl and Bi background events based either on the kinematics of the decay processes 
 or on delayed coincidence timing of the progenitors and daughter members of the natural decay chains. Although
there has been great success in reducing backgrounds in this way, all these techniques have some inefficiency. Therefore
it is necessary to minimize these activities. The future proposals will make great efforts to reduce the amount of Tl and Bi present
even if they rely on such tagging techniques.

Many isotopes not normally found in nature ({\it e.g.} $^{239,240}$Pu, $^{137}$Cs, $^{90}$Sr, $^{42}$Ar, and $^{85}$Kr are 
produced artificially by human activities such as nuclear-weapon testing, nuclear accidents, reactor ventings, etc. Therefore it is
necessary for experimenters to consider such exotic possibilities when designing an experiment.

\subsection{Cosmogenic Radioactivity}
Cosmic rays react with a detector and produce signals. Because the cosmic ray flux is so high on the surface of the 
Earth, \BB\ experiments are conducted underground. Any prompt events can be eliminated by
going to a deep location and incorporating an anti-coincidence shield. But in addition to prompt interactions,
 cosmic rays can produce delayed radioactivity via many nuclear reactions. In particular, while detector materials
or the source resides on the surface of the Earth, they are exposed to a significant fast ($>$10 MeV) neutron flux. These fast neutrons
can produce large $\Delta$A transitions in nuclei that result in radioactive nuclides.

Below ground the fast neutron flux is proportional to the cosmic-ray muon flux, so going deeper reduces it. For most cases,
a few hundred meters will suffice to eliminate the {\it in situ} production and only the residual activity left over from the time spent on 
the surface will be present. The most famous example of this effect is that of $^{68}$Ge in Ge detectors.
 Even though the experiments
used Ge enriched in $^{76}$Ge, $^{68}$Ge was produced in the crystals through the
 high-threshold reaction, $^{76}$Ge(n,9n)$^{68}$Ge.
In using enriched Ge with little $^{70,72,73,74}$Ge, experimenters had thought that the $^{68}$Ge problem would not be present because the required 
reaction on $^{76}$Ge had such a large $\Delta$A. Although it was significantly
decreased, it remained a source of background that will require additional efforts to minimize. 

For future experiments that will require sensitivities near 1 event/year in the region of interest in a 1-ton sample, the 
cosmogenic background possibilities are varied. Because the signal rate is very low in a large target, rare 
 processes must be considered as potential backgrounds. For example $^{92}$Mo(n,2n)$^{91}$Mo 
 would need to be carefully considered for any $^{100}$Mo
experiment using natural Mo. For a $^{136}$Xe experiment trying to identify the $^{136}$Ba daughter, the two-step 
process $^{136}$Xe($\mu$,p)X and $^{136}$Xe(p,n)$^{136}$Cs produces the 13.1-day Cs isotope. $^{136}$Cs decays via a 2.55-MeV
$\beta$ to $^{136}$Ba. Although it is unlikely that the energies of the $\beta$ and the coincident $\gamma$s would sum to the 
region of interest, it points out how rare (and sometimes bizarre) processes must be considered.

\section{An Ideal Experiment}
The first direct measurement of \BBt\ used a time projection chamber\cite{ELL87}. This
 was a fairly large apparatus ($\approx$ m$^3$) for a modest amount
of source (13 g) and therefore, it is doubtful that this type of arrangement will represent 
the best of the next generation of \BBz\ experiments. 
This design doesn't scale easily to very large source mass with very low backgrounds. It 
is interesting to try to enumerate the features
that an ideal \BBz\ experiment would posses. It would have the following characteristics:

\begin{itemlist}
 \item The detector mass must be large enough to reach our 50-meV goal ($\approx$ 1 ton of isotope).
 \item The \BBz\ source must be extremely low in radioactive contamination.
 \item The proposal must be based on a demonstrated technology for the detection of \BB.
 \item Although the use of natural isotope will be less costly, the enrichment process provides a good level of purification
 and also results in a (usually) much smaller volume detector. 

 \item A small detector volume minimizes internal backgrounds, which scale with the detector volume. It also minimizes external
backgrounds by minimizing the shield volume for a given stopping power. This is most easily accomplished by an apparatus 
whose source is also the detector. Alternatively, a very large source may have some advantage due to self shielding.

 \item Good energy resolution is required to prevent the tail of the \BBt\ spectrum extending 
 into the \BBz\ region of interest.
 
 \item Ease of operation is required because these experiments usually operate in remote locations.
 \item A large Q value results in a fast \BBz\ rate and also places the region of interest above many potential backgrounds.
 \item A relatively slow \BBt\ rate also helps control this background.
 \item Identifying the daughter in coincidence with the \BB\ decay energy would eliminate most potential backgrounds except \BBt.
 \item Event reconstruction, providing kinematic data such as opening angle and individual electron energy,
can aid in the elimination of backgrounds. This data might also help elucidate the physics if a statistical sample of \BBz\ events 
are observed.
 \item Good spatial resolution and timing information can help reject background processes.

 \item The nuclear theory is better understood in some isotopes than others. The interpretation 
 of limits or signals might be easier
to interpret for some isotopes.  

\end{itemlist}

No experiment, past or proposed, is able to optimize for all of these characteristics simultaneously. Each has chosen a design
that emphasizes different aspects of this list. Similar points concerning an ideal experiment
 were made in Ref. ~\refcite{ELL02} and also by Yu. Zdesenko in Ref~\refcite{ZDES}.

\section{Past Experiments}

The best experiments to date are the Ge experiments. The IGEX experiment\cite{AAL99} and the Heidelberg-Moscow experiment\cite{KLA01a}
have produced the most restrictive limits on the half-life and deduced \mee. The results from these two experiments along with
many others are given in Table \ref{Pastnu}. From the Ge experiments the limit on \mee\ is about 300 to 1300 meV, depending on the 
chosen matrix elements.

One intriguing bit of history helps motivate the need for more than one \BBz\ effort. An early spectrum in a Te experiment\cite{TePeak}
saw a peak at the 2.53-MeV endpoint with  $\approx$2$\sigma$ confidence. At a similar point in time, a Ge spectrum\cite{REU92} showed a peak of unknown
origin
at the same energy. If the Ge peak was real, it would mean that a dangerous unidentified $\gamma$ ray existed 
that would be a background for any $^{130}$Te experiment.
However, both peaks turned out to be spurious. The Te peak faded with additional statistics and the Ge peak was an electronic
artifact. The moral of the story is that a peak, by itself, may not suffice to prove an observation of \BBz. But two peaks in different
isotopes from experiments using different techniques would be convincing. This would be especially true if the rates were consistent with
a common \mee.

\begin{table}[ht]
\tbl{\protect Best reported limits on \Tz. 
The \mnu\ limits and ranges are those deduced by the authors and
their choices of matrix elements within the 
cited experimental papers. All are quoted at the 90\% confidence level 
except as noted.}
{\begin{tabular}{@{}lllc@{}} \toprule

Isotope            & T$_{1/2}^{0\nu}$ (y)                       &$\langle m_{\beta\beta}\rangle$ (eV)     & Reference \\ \colrule
$^{48}$Ca          & $>9.5\times 10^{21} (76\%)$                & $<8.3$                           & \cite{YOU91} \\
$^{76}$Ge          & $>1.9\times 10^{25}$                       & $<0.35$                          & \cite{KLA01a} \\
                   & $>1.6\times 10^{25}$                       & $<0.33-1.35$                     & \cite{AAL99}  \\
$^{82}$Se          & $>2.7\times 10^{22}(68\%)$                 & $<5$                             & \cite{ELL92}  \\
$^{100}$Mo         & $>5.5\times 10^{22}$                       & $<2.1$                           & \cite{EJI96}  \\
$^{116}$Cd         & $>7\times 10^{22}$                         & $<2.6$                           & \cite{DAN00}  \\
$^{128,130}$Te     & $\frac{T_{1/2}(130)}{T_{1/2}(128)} = (3.52\pm 0.11)\times 10^{-4}$  & $<1.1-1.5$ & \cite{BER93} \\
                   &(geochemical)                               &                                  &                 \\
$^{128}$Te         & $>7.7\times 10^{24}$                       & $<1.1-1.5$                       & \cite{BER93} \\
$^{130}$Te         & $>1.4\times 10^{23}$                       & $<1.1-2.6$                       & \cite{ALE00}\\
$^{136}$Xe         & $>4.4\times 10^{23}$                       & $<1.8-5.2$                       & \cite{LUE98} \\ 
$^{150}$Nd         & $>1.2\times 10^{21}$                       & $<3$                             & \cite{DES97} \\ \botrule
\end{tabular}}
\label{Pastnu}
\end{table}

A recent claim for \BBz\ was made by a small subset of the Heidelberg-Moscow collaboration\cite{KLA01c,KLA02}. This claim has generated
a fair amount of controversy\cite{AAL02,KLA02d,HAR02,FER02,ZDE02}. The analysis in Ref.~\refcite{KLA02} chooses a background model that contains many
peaks in an extended region and therefore limits its analysis to a narrow region about the \BBz\ peak. This analysis finds a probability 
that a peak exists at the \BBz\ location is 96.5\% or 2.1$\sigma$. A wide-region analysis assuming that only a continuous background is 
present, finds  a probability of only $\approx$65\% that a peak exists. These two extremes in the background model 
indicate that the significance of
the result is sensitive to the background description. The quoted significance of the result (2.1$\sigma$), however, does not
include a systematic uncertainty related to the choice of background model. If it did, the significance would decrease. Taken
at face value, a 
2.1$\sigma$ result may be intriguing, but it certainly does not warrant a strong claim for evidence.

\section{The Various Proposals}
The various proposals are too numerous to describe all in detail. I have listed all the proposals with which I am familiar 
in Table \ref{nufut}. To give the reader a feel for the proposals, I have chosen 5 to profile: CUORE, EXO, GENIUS, Majorana, and
MOON. The important point to take away from Table \ref{nufut} is that many of the proposals claim a sensitivity to the
desired target of 50 meV (i.e. $\approx 10^{27}$y). For many of the proposals described in Table \ref{nufut}, the previous experiments 
listed in Table \ref{Pastnu} are effective
prototypes. This fact helps build confidence that the next generation of experiments will reach their intended goal.

\begin{table}[ht]
\tbl{A summary of the double-beta decay proposals. The quoted sensitivities are those quoted by the proposers but scaled for 5 years
of run time. These sensitivities should be used carefully as they depend on background estimates for experiments that don't exist yet.}
{\begin{tabular}{@{}lclc@{}} \toprule
                           &            &                                                                   &  Sensitivity to         \\  
Experiment                 &	Source     & Detector Description                                           & $T_{1/2}^{0\nu}$ (y)    \\ \colrule
COBRA\cite{ZUB01}          &$^{130}$Te  &	10 kg CdTe semiconductors                                       & $1 \times 10^{24}$   \\
DCBA\cite{ISH00}           &$^{150}$Nd  &	20 kg $^{enr}$Nd layers between tracking chambers               & $2 \times 10^{25}$     \\
NEMO 3\cite{NEMO3}         &$^{100}$Mo  &	10 kg of \BBz\ isotope (7 kg Mo) with tracking                  & $4 \times 10^{24}$    \\   \colrule
CAMEO\cite{BEL01}          &$^{116}$Cd  &	1 t  CdWO$_4$ crystals in liq. scint.                           & $1 \times 10^{27}$   \\
CANDLES\cite{KIS01}        &$^{48}$Ca   &	several tons of CaF$_2$ crystals in liq. scint.                 & $1 \times 10^{26}$      \\
CUORE\cite{AVI01}          &$^{130}$Te  &	750 kg TeO$_2$ bolometers                                       & $2 \times 10^{26}$    \\
EXO\cite{EXO00}            &$^{136}$Xe  &	1 t $^{enr}$Xe TPC (gas or liquid)                              & $8 \times 10^{26}$    \\
GEM\cite{ZDE01}            &$^{76}$Ge   &	1 t $^{enr}$Ge diodes in liq. nitrogen                          & $7 \times 10^{27}$      \\
GENIUS\cite{KLA01b}        &$^{76}$Ge   &	1 t 86\% $^{enr}$Ge diodes in liq. nitrogen                     & $1 \times 10^{28}$    \\
GSO\cite{DAN01,WANGS01}    &$^{160}$Gd  &	2 t Gd$_2$SiO$_5$:Ce crystal scint. in liq. scint.              & $2 \times 10^{26}$   \\
Majorana\cite{MAJ01}       &$^{76}$Ge   &	0.5 t 86\% segmented $^{enr}$Ge diodes                          & $3 \times 10^{27}$   \\  
MOON\cite{EJI00}           &$^{100}$Mo  &	34 t $^{nat}$Mo sheets between plastic scint.                   & $1 \times 10^{27}$   \\   
Xe\cite{CAC01}             &$^{136}$Xe  & 1.56 t of $^{enr}$Xe in liq. scint.                               & $5 \times 10^{26}$    \\
XMASS\cite{XMASS}          &$^{136}$Xe  &  10 t of liq. Xe                                                  & $3 \times 10^{26}$    \\ \botrule
\end{tabular}}

\label{nufut}
\end{table}

\subsection{CUORE}
The success of the MIBETA experiment\cite{ALE00} has resulted in the CUORE (Cryogenic Underground Observatory for 
Rare Events) proposal\cite{PIR00}. One thousand TeO$_2$ crystals of 750 g each 
would be operated as a collection of
bolometers. The detectors will be collected into 25 separate towers of 40 crystals. Each tower will 
have 10 planes of 4 crystals each. One such plane has already been successfully tested and a single 
tower prototype referred to as CUORICINO is scheduled to begin operation in spring 2003. 

The energy resolution at the \BBz\ peak (2.529 MeV) is expected to be about 5 keV FWHM 
($\approx$ 0.2\%). The 
background has been measured in the first plane to be 
$\approx$0.5 counts/(keV$\cdot$kg$\cdot$y). However a major 
component of this background was due to a surface contamination arising from the use of cerium 
oxide polishing compound which tends to be high in thorium. With this problem solved,
the experimenters project a 
conservative estimate of the background to be $\approx$0.01 counts/(keV$\cdot$kg$\cdot$y).

A major advantage of this proposal is that the natural abundance of $^{130}$Te is 34\% and, thus, 
no enrichment is needed resulting in significant cost savings. As with MIBETA, the cosmogenic
activities within the TeO$_2$ crystals are not a serious concern. On the other hand, the crystal
 mounts and cryostat form a significant amount of material close to the bolometers. Much of the 
cryostat is shielded with Roman period lead but a fair quantity of copper and Teflon remain close 
to the crystals.

\subsection{EXO}
The Enriched Xenon Observatory (EXO)\cite{EXO00} 
proposes to use up to 10 t of 60-80\% enriched $^{136}$Xe.
The unique aspect of this proposal is the plan to detect 
the $^{136}$Ba daughter ion correlated with the
decay. If the technique is perfected, it would eliminate 
all background except that associated with \BBt. 
The real-time optical detection of the daughter Ba ion, initially suggested in Ref.~\refcite{MOE91},  might 
be possible if the ion can be localized and probed with lasers. The spectroscopy has been used for 
Ba$^+$ ions in atom traps. However, the additional technology to 
detect single Ba ions in a condensed medium 
or to extract single Ba ions from a condensed medium and trap them must be 
demonstrated for this application. To optically detect the alkali-like Ba$^+$ ion, it is excited 
from a 6$^2$S$_{1/2}$ ground state to a 6$^2$P$_{1/2}$ with a 493-nm laser. Since this excited state
 has a 30\% branching ratio to a 5$^4$D$_{3/2}$ metastable state, the ion is detected by re-exciting
 this metastable state to the 6P state via a 650-nm laser and then observing the resulting decay back to the
 ground state. This procedure can be repeated millions of times per second on a single ion and produce a significant signal.

The EXO plan is to use Liquid Xe (LXe) 
scintillator. The LXe concept has the advantage of being much smaller than a gaseous TPC due to the high density
 of LXe. However, the higher density makes the scattering of the laser light too great to optically detect
 the Ba$^+$ {\it in-situ}. However, once the Ba ion is localized via its scintillation 
and ionization, it might be extracted 
via a cold finger electrode coated in frozen Xe (M. Vient, unpublished observation, 1991). 
The ion is electrostatically attracted to
 the cold finger which later can be heated to evaporate the Xe and release the Ba ion into a radio 
frequency quadrupole trap. At that point, the Ba$^{++}$ is neutralized to Ba$^+$, laser cooled and
 optically detected. The efficiency of the tagging has yet to be demonstrated and is a focus of current
research.

The collaboration has recently performed experiments to optimize the energy resolution.
By measuring both scintillation light and ionization simultaneously, they have achieved energy resolution
sufficient for the experiment. Tests to determine the viability of the Ba
 extraction  process are also being performed. 
The EXO collaboration has received funding to proceed with a 200-kg enriched Xe
detector without Ba tagging. This initial prototype will operate at the Waste Isolation Pilot Plant
(WIPP) in southern New Mexico.

\subsection{GENIUS}
The progress and understanding of Ge detectors has been developed over more than 30 years 
of experience. The potential of these detectors lie in their great energy resolution, ease of 
operation, and the extensive body of experience relating to the reduction of backgrounds. 
This potential is not yet exhausted as is evidenced by the GENIUS and Majorana proposals 
that build on the experimenters' previous efforts.

The GENIUS (GErmanium NItrogen Underground Setup)\cite{KLA01b}
 proposal has evolved from the Heidelberg-Moscow (HM) experiment.
 The driving design principle behind this proposed Ge detector array experiment is the 
evidence that the dominant background in the HM experiment was due to radioactivity external
 to the Ge. (The reader should contrast this with the motivation for the design of the 
Majorana proposal described below.) An array of 2.5-kg, p-type Ge crystals would be operated
 ''naked" within a large liquid nitrogen (LN) bath. By using naked crystals, the external 
activity would be moved to outside the LN region. P-type crystals have a dead layer on the
 external surface that reduces their sensitivity to external $\beta$ and $\alpha$ activity.
 Due to its low stopping power, roughly 12 m of LN is required to shield the crystals from 
the ambient $\gamma$-ray flux at the intended experimental site at 
Gran Sasso. By immersion in LN, the optimal operating temperature 
is maintained without a bulky cryostat and a test of the naked operation of a crystal in a 50 l
 dewar has been successful\cite{KLA98,BAU99}. The results from a several-week operation indicate that the performance of the 
detector was comparable to those operated in a conventional vacuum-tight cryostat system. 
Their measurements also indicate very little cross talk between naked detectors and that 
an extended distance ($\approx$6 m) between the FET and the crystal does not degrade the energy resolution.

The proposal anticipates an energy resolution of $\approx 6$ keV FWHM (0.3\%) and a threshold of 11 keV. The value of
 this low threshold is set by x rays from cosmogenic activities. Using 1 t of 86\% enriched 
Ge detectors, the target mass is large enough for dark matter
 studies. In fact a 40-kg $^{nat}$Ge proof-of-principle experiment has been approved for dark
 matter studies.

\subsection{Majorana} The Majorana proposal\cite{MAJ01} (named in honor of Etorre Majorana) involves many of the IGEX 
collaborators. Their analysis indicated that $^{68}$Ge contained within the Ge detectors
 was the limiting background for their \BBz\ search. (Contrast this with the GENIUS approach
described above.) The proposal's design therefore 
emphasizes segmentation and pulse shape discrimination to reject this background. The
 electron capture of $^{68}$Ge is not a significant problem but $^{68}$Ge decays
 to the $\beta^+$ emitting $^{68}$Ga. This isotope can create background in the \BBz\ 
window if one of the annihilation $\gamma$ rays converts within the crystal. The energy 
deposits of the positron and $\gamma$ ray may pollute the peak window in energy, but the 
deposits will be separated in space. In contrast, a \BBz\ event will have a localized 
energy deposit. Segmentation of the crystals permits a veto of such events. Furthermore,
 distinct ionization events will have a different pulse shape than a localized event. 
Therefore pulse shape analysis can also help reject background. 
Majorana plans to use $\approx$ 500, 86\% enriched, segmented Ge crystals for a total of 500 kg of
 detector. The cryostat would be formed from very pure electroformed Cu\cite{BRO90}.

\subsection{MOON}

The MOON (Mo Observatory Of Neutrinos) proposal\cite{EJI00} plans to use $^{100}$Mo as a \BBz\ source
 and as a target for solar neutrinos. This dual purpose and a sensitivity to low-energy supernova electron
neutrinos\cite{EJI01}
 make it an enticing idea. $^{100}$Mo has a high Q-value (3.034 MeV), which results in a large
phase space factor and places the \BBz\ peak region well above most radioactive backgrounds. It also 
has hints of a favorable
 \Mz\ but unfortunately it has a fast \Tt. The experiment will make energy and
angular correlation studies of \BB\ to select \BBz\ events and to
reject backgrounds. The planned MOON configuration is a supermodule of scintillator and Mo ensembles. One
option is a module of plastic fiber scintillators with thin (0.03 g/cm$^2$)
layers of claded Mo, which are arranged to achieve a position resolution 
comparable to the fiber diameter (2-3 mm). A total of 34 tons of natural Mo would be required. 

As a solar neutrino detector, $^{100}$Mo has a low threshold: 168 keV, and the estimated 
observed event rate is $\approx$160/(ton $^{100}$Mo$\cdot$year) without neutrino oscillations. 
It is sobering to realize
 that the primary background for the delayed-coincidence solar neutrino signal
 is accidental coincidences between \BBt\ decays. 

The project needs Mo and scintillator radioactive impurity levels of better than 1 mBq/ton.
 This can be achieved by carbonyl chemistry for Mo and plastics can be produced cleanly. 
However, the total surface area of the Mo-scintillator modules is $\approx$26000 m$^2$.
 Dust, being electrostatically charged, tends to garner Rn daughters and 
becomes radioactive. Keeping these surfaces clean of dust during production and assembly will be a challenge.
 Liquid scintillator and bolometer options that would avoid this large surface area are also being considered. 
The simulations of the design indicate that the energy resolution for the \BBz\ peak will be
 $\approx$7\% which is at the upper end of the range of feasibility for a sub 50 meV \mnu\ experiment.
The bolometer option would also remove the resolution concerns.
Use of enriched  $^{100}$Mo is feasible, and would reduce
the total volume of the detector and source ensemble resulting in a lower
internal radioactivity contribution to the background by an order of magnitude.

\subsection{OTHER PROPOSALS}

There are too many proposals for detailed description so I have summarized those
of which I am aware in Table \ref{nufut} and mention them here. The CAMEO proposal\cite{BEL01}
 would use 1000 kg of scintillating $^{116}$CdWO$_4$
crystals situated within the Borexino apparatus. The Borexino liquid scintillator would provide
shielding from external radioactivity and light piping of crystal events to 
the photomultiplier tube (PMT) array surrounding
the system. Early phases of the program would use the Borexino counting test facility. Similarly, 
the CANDLES proposal\cite{KIS01} (CAlcium floride for study of Neutrino and Dark matter by Low Energy Spectrometer) plans to 
immerse CaF$_2$ in liquid scintillator. The scintillation light from the $\beta\beta$ of $^{48}$Ca will be detected
via PMTs. The low isotopic abundance (0.187\%) of $^{48}$Ca requires a very
large operating mass. Two groups\cite{DAN01,WANGS01} have been studying the use of  GSO crystals (Gd$_2$SiO$_5$:Ce) for the study of
$^{160}$Gd. 

COBRA (CdTe O neutrino double Beta Research Apparatus)\cite{ZUB01} 
would use CdTe or CdZnTe semiconductors to search for \BBz\ in either Cd or Te. 1600 1-cm$^3$ 
crystals would provide 10 kg of material. GEM is a proposal\cite{ZDE01} that is very similar to that
of GENIUS. However, much of the LN shielding would be  replaced with high purity
water.

The Drift Chamber Beta-ray Analyzer (DCBA) proposal\cite{ISH00} is for a three-dimensional tracking chamber in a 
uniform magnetic field. A drift chamber inside a solenoid and cosmic-ray veto counters comprises the detector.
Thin plates of Nd would form the source. The series of NEMO experiments is progressing with NEMO-3\cite{NEMO3}
beginning operation in 2002. In concept the detector is similar to NEMO-2. That is, it 
contains a source foil enclosed between tracking chambers that is itself enclosed within a scintillator
array. NEMO-3 can contain a total of 10 kg of source and plans to operate 
with several different isotopes,
but with $^{100}$Mo being the most massive at 7 kg. The collaboration is also discussing 
the possibility
of building a 100-kg experiment that would be called NEMO-4.

There are two additional groups proposing to use $^{136}$Xe to study \BBz. Caccianiga and Giammarchi\cite{CAC01} propose to 
dissolve 1.56 t of enriched Xe in liquid scintillator. The XMASS\cite{XMASS} collaboration proposes to use
10 t of liquid xenon for solar neutrino studies. The detector would have sensitivity to \BBz.

\section{The Matrix Elements}
In order to determine an effective Majorana neutrino mass from a \BBz\ result, one requires input from
nuclear theory. The matrix elements however are very hard to calculate and the various techniques
and authors produce varying results. Reference \refcite{SUH98} includes a rather complete list of references and the 
results of calculation of the matrix elements. The two techniques for calculating the matrix elements are
the quasiparticle random phase approximation (QRPA) and the nuclear shell model. Each approach has its advantages
and disadvantages and hence critics and advocates. Due to this complicated situation, the spread in the
calculations is commonly taken as a measure of the theoretical uncertainty.

For a given \mee, the spread in predicted half-lives is approximately an order of magnitude.
Thus the spread in \mee\ deduced from a potential \BBz\ measurement is a factor of 3. It should be emphasized that
the qualitative physics conclusions arising from an observation of \BBz\ are so profound, that this factor of
3 is a detail. However, precise interpretation of any result will be limited due to this spread.
The spread is due to the community's choice to include {\it all} calculations when estimating the
theoretical uncertainty. No consensus has been reached identifying the calculations that are outdated or relatively less
precise. Naively, one concludes that the factor of 3 is an overestimate. At present this uncertainty is large enough that
potential conclusions regarding the mass spectrum\cite{ELL02} or the Majorana phases\cite{BAR02} might be diminished.
Improvement in the calculations will have a big payoff if \BBz\ is observed.

More important, however, is the need for confirmation as described above. If observations of \BBz\ are made in two isotopes,
a comparison between the two deduced \mee\ values might indicate that they are consistent. Obviously this would be
strong evidence that \BBz\ is truly the correct interpretation. For this
argument to be persuasive, however, confidence in the theory and its estimated uncertainties must be sufficient that critics
concede the two \mee\ values are consistent.

Encouragingly, the technology available for such calculations has improved, especially the shell model techniques. The physics
motivation for these experiments is very strong, so perhaps interest in the theoretical aspect of this problem will
build also.

\section{Conclusions}
It is a very exciting time for \BBz. The next generation of experiments will have a sensitivity of about 50 meV
for \mee: the critical target of $\sqrt{\delta m_{atm}^2}$. Therefore future experiments will have
a known neutrino mass with which to compare. As a result, even null experiments will impact the question
of whether neutrinos are Dirac or Majorana in certain mass-spectrum configurations.

\section*{Acknowledgments}
After the announcement of the 2002 Nobel Prize for physics, this conference 
evolved into a tribute to Ray Davis and 
his great achievements that culminated with his winning this prestigious award. I 
wish to join the numerous scientists to congratulate him on a spectacular career.
 I had the pleasure to collaborate with Ray on the Russian-American
Gallium solar neutrino Experiment (SAGE) and, therefore, 
I also wish to thank him personally for his inspiring example.

I wish to thank Frank Avignone and Peter Doe for critical readings of this manuscript.
This research was sponsored in part by DOE grant W-7405-ENG-36.

\end{document}